\documentclass[apl,twocolumn,reprint]{revtex4}
\usepackage{bm}
\usepackage{graphicx}
\usepackage{color}

\begin{document}


\title{
\textcolor{blue}{Control of Circularly Polarized THz Wave from Intrinsic Josephson Junctions by Local Heating}
}

\author{Hidehiro Asai and Shiro Kawabata}
\affiliation{Nanoelectronics Research Institute (NeRI), National Institute of Advanced Industrial Science and Technology (AIST), Tsukuba, Ibaraki 305-8568, Japan}
\date{\today}
\begin{abstract}
This paper reports a practical method of generating circularly polarized terahertz (THz) waves from intrinsic Josephson junctions (IJJs) 
and controlling their polarization states by external local heating.
We theoretically find that a mesa-structure IJJ whose geometry is almost square can emit circularly polarized THz waves by local heating of the mesa.
Moreover, we demonstrate that the polarization states of the THz waves change dramatically with the local heating position.
Our results indicate that the use of local heating can provide a high level of controllability of the THz emissions
and significantly extend the range of applications of IJJ-based THz emitters.
\end{abstract}

\pacs{}
\maketitle 

Terahertz (THz) electromagnetic (EM) waves have potential applications in a wide range of fields,
 e.g.  non-destructive inspection of materials, medical diagnosis, bio-sensing, and high-speed wireless communication~\cite{Tonouchi}.
Toward the realization of a practical THz emitter, various THz sources, such as quantum-cascade lasers and resonant-tunneling diodes, have been studied so far~\cite{qcl,tdiode}.
Since the first observation of intense and continuous THz wave emitted from a $\textrm{Bi}_2\textrm{Sr}_2\textrm{CaCu}_2\textrm{O}_{8+\delta}$ (Bi2212) single crystal by
 Ozyuzer $et$ $al$.~\cite{Ozyuzer}, considerable attention has been paid to
 high-$T_{\rm c}$ cuprate superconductors.   
A layered structure of Bi2212 single crystals composed of superconducting layers and insulating layers 
forms a stack of the Josephson junctions whose thicknesses are on the atomic scale $\sim 1.5$  nm.  
Stacks of these natural Josephson junctions, which are referred to as intrinsic Josephson junctions (IJJs), can generate an AC Josephson current in the THz frequency range by application of a voltage, and the THz wave emission is attributed to the generation of this current.
Intense THz emissions have been reported for IJJs fabricated in a mesa geometry, which itself behaves as a cavity resonator,  and thus,
 a number of studies on such IJJ mesas have been carried out both experimentally~\cite{Ozyuzer,jpsjKado,hotWang,hot2Wang,aplKakeya,tunableTsujimoto,TBenseman,An,Sekimoto,heatTsujimoto,Watanabe,Zhou}
 and theoretically~\cite{inphaseKoshelev,kinkHu1,kinkKoshelev,fdtdKoyama1,inphaseNori,Klemm,inphaseAsai,hotGross2,patchAsai,laserAsai,IEEEAsai}.
However, in the previous studies, little attention was given to the state of the polarization of the THz waves.
Control of the polarization is a fundamental issue for technological applications of THz waves.
In particular, the ability to generate and control circularly polarized THz waves would open the way to various applications such as circular dichroism of proteins~\cite{CD}.

In this paper, we present a method of generating and controlling circularly polarized THz waves from IJJ mesas by using external local heating.
We consider a IJJ mesa whose geometry is almost square and whose temperature distribution is controlled by laser heating (see Fig. 1).
We numerically investigate the emission power and the polarization of the THz waves from the mesa by solving the sine-Gordon and the Maxwell equations simultaneously. 
Interestingly, both the ellipticity and handedness of the THz waves from the IJJ mesa strongly depend on the position of the laser heating.

\begin{figure}[htbp]
\begin{center}
\includegraphics[width = 8.5 cm]{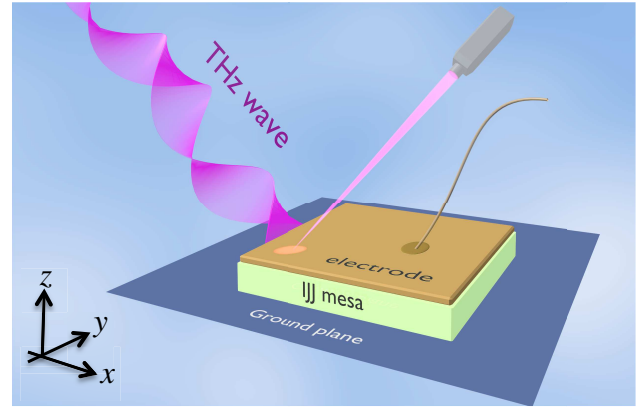}
\caption{Schematic figure of IJJ mesa whose geometry is almost square, and is locally heated by laser irradiation.}
\end{center}
\label{f1}
\end{figure}

Figure 1 shows a schematic figure of THz emissions from an IJJ mesa whose polarization is controlled by laser heating.
 A DC voltage is applied to the mesa, whose width $W$ and the length $L$ of the mesa are almost equal,
 and the temperature distribution of the mesa is controlled by a focused laser beam.
Firstly, the resonant frequencies of the cavity modes along the $x$-axis, TM(1,0), and the $y$-axis, TM(0,1), are close to each other 
in this mesa because $W$ and $L$ are almost equal.
Mixture of these modes with a $\pm \pi/2$ phase difference leads to the generation of circularly polarized THz waves.
Secondly, the local increase of the mesa's temperature by  the laser heating results in a local change in the critical current density $j_{\textrm c}$.
 Theoretical studies have pointed out that such an inhomogeneous $j_{\textrm c}$ distribution affects the excitation of cavity modes in the mesa \cite{inphaseKoshelev,inphaseAsai}.
Moreover, the ability of laser heating to change the THz emission power dramatically has been predicted theoretically and shown experimentally \cite{laserAsai,heatTsujimoto,Watanabe,Zhou}.
 Therefore, by manipulating the $j_{\textrm c}$ distribution by laser heating, we should be able to control the intensity of the TM(1,0) and the TM(0,1) modes in IJJ mesa and thus control the state of the circular polarization.

 To verify this idea for polarization control, 
 we performed a numerical simulation on THz emission from a IJJ mesa such as the one described above. 
 On the basis of the antenna theory, we set $W$ and $L$ so that $(L-W)/L$ almost equals to the Q factor of the emitter. 
This restriction results in the $\pm \pi/2$ phase difference between two orthogonal cavity modes when the emission frequency equals to the middle of the resonant frequencies of those modes\cite{Balanis}.
We set $W = $ 78 $\mu$m, $L =$ 80 $\mu$m, and set the thicknesses of the electrode and mesa to 2 $\mu$m.
 We used an in-phase approximation in which all phase differences are equal to a common phase difference  $\phi$,~\cite{fdtdKoyama1,inphaseKoshelev,inphaseAsai}
and the dynamics of $\phi$ is given by,  
\begin{eqnarray}
\frac{\hbar \epsilon_c}{2eD} \frac{\partial^2 \phi}{\partial t^2}   =   c^2  \Bigl[ \frac{\partial B_y}{\partial x} - \frac{\partial B_x}{\partial y} \Bigr] - \frac{1}{\epsilon_0 } \left[  j_{\textrm{c}}  \sin{\phi} + \sigma_c E_z -j_{\textrm{ex}} \right] ,
\end{eqnarray}
where $\epsilon_{\textrm c}$ is the dielectric constant of the junctions, $c$ is the light velocity, $j_{\textrm{ex}}$ is the external current.
The electromagnetic (EM) fields in the IJJs are given by $E_{z} = \frac{\hbar}{2eD}  \frac{\partial \phi}{\partial t}$, $B_{y} = \frac{\hbar}{2eD}\frac{\partial \phi}{\partial x}$, $B_{x} = -\frac{\hbar}{2eD}\frac{\partial \phi}{\partial y}$
and $D$ is the thickness of the insulating layers of the IJJs.
Moreover, we set $\epsilon_{\textrm c} = 17.64$,  $D = 1.2$ nm, $\sigma_{\textrm c} = 10$ S/m, and $j_{\textrm{c}} = 2.18 \cdot 10^3$ A/${\textrm {cm}}^2$.
For simplicity, the laser heating was simulated by decreasing $j_{\textrm c}$ in a square area (20 $\mu$m $\times$ 20 $\mu$m) in the $x$-$y$ plane,~\cite{inphaseAsai}
; the heating positions $\mathcal{A} \sim \mathcal{G}$ are indicated in Fig. 2.
In this study, we assumed that the maximum temperature of the mesa  without the external heating 
is much lower than $T_{\textrm c}$ due to high thermal cooling through a metal substrate. 
 In this case, the local hot spot does not introduce a large damping of $j_{\textrm c}$. 
 On the other hand, as we reported in our previous work\cite{laserAsai}, the laser heating can decrease $j_{\textrm c}$ around the heating spot by $30\sim100$\%.
Note that 100\% decrease in $j_{\textrm c}$ indicates the temperature of the spots exceed the $T_{\textrm c}$. 
In this previous calculation, we assumed a Bi2212 substrate whose thermal conductivity is low, 
and the small heat flow to the substrate is attributed to the increase in the local temperature up to $T_{\textrm c}$.
Meanwhile, in this study, we assumed a normal metal substrate, 
and the high thermal cooling through the substrate will keep the mesa temperature lower than $T_{\textrm c}$. 
In this study, we considered 50\% decrease in $j_{\textrm c}$ around the heating spot as an example of the moderate change of $j_{\textrm c}$ by the local heating.

\begin{figure}[htbp]
\begin{center}
\includegraphics[width = 5.5 cm]{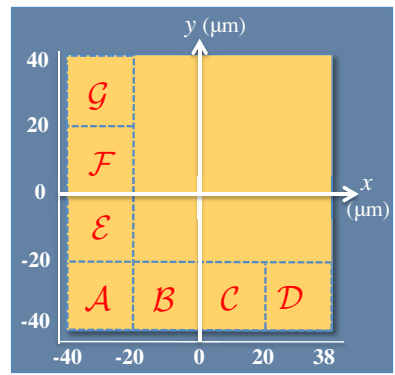}
\caption{Positions of the heating spot on the mesa $\mathcal{A} \sim \mathcal{G}$ in the $x$-$y$ plane.}
\end{center}
\label{f2}
\end{figure}

As mentioned before, the excitation of the TM(1,0) and TM(0,1) modes in the mesa must be excited in order to generate circularly polarized waves.
The electric fields of the TM(1,0) and TM(0,1) modes are given by, $E_{1,0} = A_{1,0} \sin{(\pi x/W)}, E_{0,1} = A_{0,1} \sin{(\pi y/L)}$, 
where $A_{1,0}$ and $A_{0,1}$ are the amplitudes of the TM(1,0) and TM(0,1) modes, respectively. 
The resonant frequencies of the cavity modes are given by 
$f_{1,0} = c/(2 \hspace{-0.8mm}\sqrt{\epsilon_c}W) = 0.458$ THz  
for the TM(1,0) mode,
and 
$f_{0,1} = c/(2 \hspace{-0.8mm}\sqrt{\epsilon_c}L) = 0.446$ THz
for the TM(0,1) mode.
The linear combination of these cavity modes yields the mixed electromagnetic mode, $E_m$, in the mesa, 
\begin{eqnarray}
E_m (x,y,t) = \cos{(2\pi f_{i} t + \delta_s)} E_{1,0}+ \cos{(2\pi f_{i} t )} E_{0,1}. \label{cavity}
\end{eqnarray}
 Here, $f_{i}$ is the frequency of the input wave (i.e. AC Josephson frequency), $\delta_s$ is the phase differences between the TM(1,0) and TM(0,1) modes.
As described in these equations, the TM(1,0) mode is antisymmetric with respect to the $y$ axis, while the TM(0,1) mode is antisymmetric with respect to the $x$-axis. 
Hence, the $j_{\textrm{c}}$ distribution must be antisymmetric with respect to both axes 
 in order to excite both cavity modes.~\cite{inphaseAsai,inphaseKoshelev} 
\begin{figure}
\begin{center}
\includegraphics[width = 6.5cm]{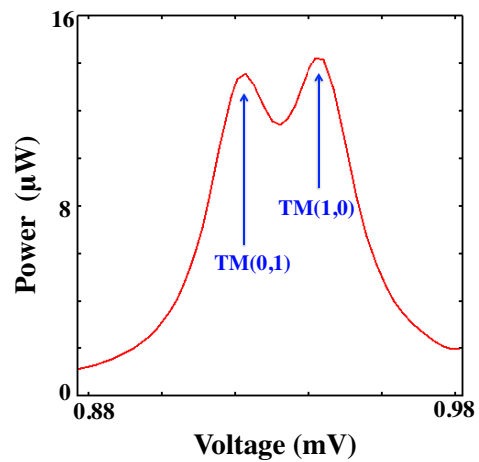}
\end{center}
\caption{Emission power vs. voltage curve as calculated for the heating position $\mathcal{A}$. The two peaks indicated by the arrows correspond to the excitation of the TM(0,1) mode and TM(1,0) modes, respectively.}
\label{f3}
\end{figure}
Thus, to begin with, we examined the THz emissions from the mesa for heating position $\mathcal{A}$, which is at the corner of the mesa.
We calculated the THz emission power and far field from the mesa around the voltages at which the frequencies of the emission waves 
satisfy the cavity resonant conditions for the TM(1,0) and the TM(0,1) modes. 
 The far fields along the $z$-axis were calculated from the equivalent electric and magnetic current along the surface of the calculation region~\cite{FDTDfar}.
Figure 3 indicates emission power as a function of voltage per IJJ layer for the heating position $\mathcal{A}$.
As can been seen from this figure, two emission peaks appear around 0.923 mV and 0.941mV, 
and they correspond to the excitation of the TM(1,0) mode and the TM(0,1) mode, respectively.
The peak emission powers are $\sim 10 \mu$W, which is comparable to those of the experimental studies~\cite{Sekimoto}.
Note that the two peaks heavily overlap because the resonant frequencies of the cavity modes are close to each other.
Thus, the emission power at the middle of the two peak voltages 0.932 mV, at which the TM(1,0) and TM(1,0) modes mix, is also high $10 \mu$W.
\begin{figure}
\includegraphics[width = 8.5cm]{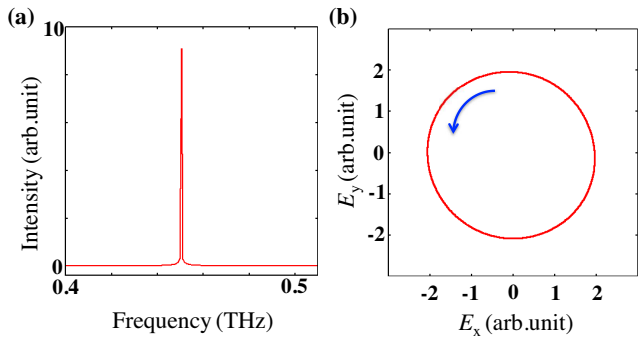}
\caption{(a) Frequency spectrum of THz waves emitted from IJJ mesa, 
and (b) trajectory of the far electric field along the $z$-axis for the heating position $\mathcal{A}$. 
The blue arrow indicates the time evolution of the electric field. }
\label{f4}
\end{figure}
Figure 4 (a) shows the frequency spectrum of the EM wave from the IJJ mesa at 0.932 mV.
As we reported in our previous work, we can see that the peak frequency of the emission is 0.451 THz, according to the AC Josephson relation $f_J = 2eV/h$, 
and the emission frequency is almost the average of the resonant frequencies of the TM(1,0) mode and TM(0,1) mode\cite{IEEEAsai}.
Figure 4 (b) is the trajectory of the far electric field along the $z$-axis at 0.932 mV.
The arrow in the figure is the direction of the time evolution.
As we expected, the emissions from both modes mix with $| \delta_s | =\pi/2$,
and circularly polarized waves are certainly observed.
%
%

%
%
Finally, to clarify the dependence of the state of the polarization on the heating position, 
we investigated the THz emissions from the IJJ mesa by varying the heating position. 
Similar to the above calculation, we calculated the far electric fields at 0.932 mV for each heating position.
Figures 5 (a)$\sim$(f) show the trajectories of the far electric field along the $z$-axis for the heating positions $\mathcal{B}\sim \mathcal{G}$, respectively.
In these figures, the ellipticity and the handedness of the polarization significantly change with the heating position.
Of particular note is that the tilted elliptical polarizations were not observed in this calculation. This is because we fixed the bias voltage so that the AC Josephson frequency equals to $(f_{1,0} + f_{0,1})/2$. In this condition $| \delta_s | =\pi/2$ is always satisfied, and the tilted polarization never appear. However, as reported in our previous work\cite{IEEEAsai}, we can generate tilted elliptical polarization by sweeping the bias voltage which causes the change of $| \delta_s |$ from $\pi/2$.
In this case, the change in ellipticity can be understood from the difference between the excitation intensities of the cavity modes.
As we reported in our previous work, the strong excitation of the anti-symmetric cavity modes such as TM(1,0) and the TM(0,1) mode
occurs when the edges of the mesa are heated~\cite{laserAsai}.
Heating position $\mathcal{B}$, for example, is the edge with respect to the $y$-axis, but not the 
edge with respect to the $x$-axis.
Thus, the excitation of the TM(0,1) mode becomes stronger than that of the TM(1,0) mode (i.e., $|A_{0,1}| > |A_{1,0}|$),
and this results in the appearance of an elliptical polarization whose minor axis is the $x$-axis, as shown in Fig. 5 (a) .  
On the other hand, heating position $\mathcal{A}$ is the edge with respect to the both the $x$ and $y$-axes.
Hence, the excitation intensities of the TM(0,1) and TM(1,0) modes are equal (i.e., $|A_{0,1}| \approx |A_{1,0}|$), and thus, almost perfectly circularly polarized waves appear.
In addition, the change in the handedness can be explained by the difference in the phases of the cavity modes.
Regarding the heating spots $\mathcal{A}$ and $\mathcal{D}$, for example, the $j_c$ distributions along the $y$-axis are same.
However, the $j_c$ distributions along the $x$-axis are antisymmetric with respect to the $y$-axis.
 Since the TM(1,0) mode is antisymmetric with respect to the $y$-axis,
mode excitations by this antisymmetric perturbation of $j_c$ results in a difference in sign of the mode amplitudes, $A_{1,0}$.
The difference in sign of the amplitude indicates a $\pi$ difference in the phase of the cavity mode,
and thus, the handedness of the circular polarization reverses.
A similar arguments, of course, holds for the TM(0,1) mode, and the reversal of handedness also occurs for the case of heating positions $\mathcal{A}$ and $\mathcal{G}$.
A sequence of the snapshots of the electromagnetic modes in the mesa for the heating spots $\mathcal{A}$, $\mathcal{B}$ and $\mathcal{D}$ are given in a supplemental material,
and they support the above discussion.
As discussed so far, the state of the polarization dramatically changes with the position of the local heating.
Thus, our results indicate that the state of the circular polarization can be controlled using local heating such as by laser irradiation.
\begin{figure}
\begin{center}
\includegraphics[width = 8.5cm]{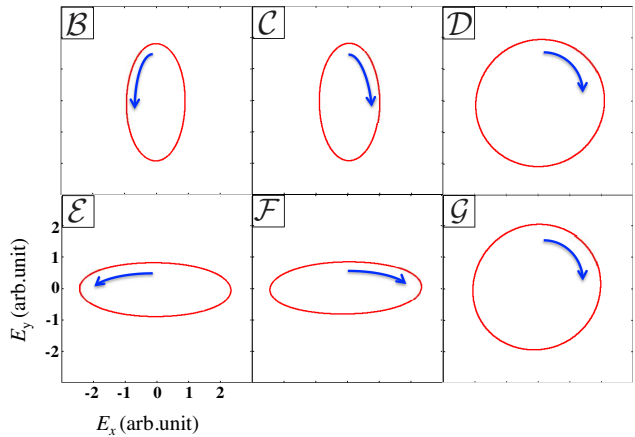}
\end{center}
\caption{(a)~(f) Trajectories of the far electric fields along the $z$ axis for the heating position $\mathcal{B} \sim \mathcal{G}$, respectively. 
Blue arrows indicate the time evolution of the electric fields.}
\label{f5}
\end{figure}

In conclusion, 
we devised a practical method of generating and controlling circularly polarized THz waves from intrinsic Josephson junctions (IJJs) 
by using external local heating.
For obtaining the circularly polarized THz waves, a mixture of two kinds of THz waves whose polarizations are orthogonal to each other is required.
In this study, we examined a mesa-structure IJJ whose width and length are almost equal and excited TM(1,0) and TM(0,1) mode simultaneously by using external local heating.
We found that the mesa emits circularly polarized THz waves when we locally heated the corners of the mesa.
Furthermore, we revealed that the ellipticity and handedness of the THz waves change dramatically with the heating position. 
Recently, emissions of circularly polarized THz wave were experimentally reported from a truncated-edge square mesa~\cite{ElarabiPlasma2016}, and their characteristics were investigated with a simple antenna analysis~\cite{ElarabiProcedia}.
In the truncated-edge mesa, the cavity modes along two diagonal lines of the mesa will become eigenmodes. Since the eigenmodes are different from our system, the dependence of the polarization on the heating spots will also change. However, similar to the rectangular mesa, the excitation of these orthogonal eigenmodes can be controlled by the heating position. Hence, we can make a similar argument on the polarization control of the THz wave from the truncated-edge mesa.
Our results add to the understanding gained in those previous studies, and more importantly,
predict that any kind of polarized THz wave can be generated from a single IJJ mesa by using external heating. 
See supplementary material for the snapshots of the electromagnetic modes in the mesa changing with the heating position. 

H.A. was partially supported by a Grant-in-Aid for JSPS
(Japan Society for the Promotion of Science) Fellows, and a
Grant-in-Aid for Young Scientists (B) from JSPS (Grant No.
26790062). S.K was partially supported by a Grant-in-Aid
for Scientific Research (C) from JSPS (Grant No. 24510146).

\end{document}